\documentclass[12pt]{article}

\usepackage{wrapfig}
\usepackage{graphicx}
\usepackage[none]{hyphenat}
\usepackage[english]{babel}
\usepackage[utf8]{inputenc}
\usepackage[T1]{fontenc}
\usepackage{makeidx}

\usepackage{authblk}
\usepackage{tikz-cd}

\usepackage{caption}
\usepackage{csquotes}
\usepackage{mathtools,amssymb,amsfonts,amsthm}

\usepackage[breaklinks,unicode=true]{hyperref}
\usepackage[capitalise]{cleveref} % \cref{…} (References)

\theoremstyle{plain}
\newtheorem{theorem}{Theorem}
\newtheorem*{theorem*}{Theorem}

\newtheorem*{lemma*}{Lemma}
\newtheorem{proposition}[theorem]{Proposition}
\newtheorem*{proposition*}{Proposition}
\newtheorem{corollary}[theorem]{Corollary}
\newtheorem*{corollary*}{Corollary}

\theoremstyle{definition}
\newtheorem{definition}{Definition}
\newtheorem*{definition*}{Definition}
\newtheorem{example}{Example}
\newtheorem*{example*}{Example}

\crefname{theorem}{Theorem}{Theorems}
\Crefname{theorem}{Theorem}{Theorems}
\crefname{lemma}{Lemma}{Lemmas}
\Crefname{lemma}{Lemma}{Lemmas}
\crefname{proposition}{Proposition}{Propositions}
\Crefname{Prop}{Proposition}{Propositions}
\crefname{corollary}{Corollary}{Corollaries}
\Crefname{corollary}{Corollary}{Corollaries}
\crefname{definition}{Definition}{Definitions}
\Crefname{definition}{Definition}{Definitions}
\crefname{example}{Example}{Examples}
\Crefname{example}{Example}{Examples}
\crefname{theorem}{Theorem}{Theorems}

\newtheorem*{exercise*}{Exercise}
\crefname{exercise}{exercise}{exercises}
\Crefname{exercise}{Exercise}{Exercises}  

\theoremstyle{remark}
\newtheorem{remarkx}{Remark}
\newtheorem*{remarkx*}{Remark}
\crefname{remark}{Remark}{Remarks}
\Crefname{remark}{Remark}{Remarks}
\newenvironment{remark}
  {\pushQED{\qed}\remarkx}
  {\popQED\endremarkx}
\newenvironment{remark*}
  {\pushQED{\qed}\remarkx*}
  {\popQED\endremarkx*}

\title{The evolution equation: an application of groupoids to material evolution}

\author[-]{V\'ictor Manuel Jim\'enez}
\author[+*]{Manuel de León}
\affil[1]{\href{mailto:victormanuel.jimenez@uah.es}{victormanuel.jimenez@uah.es}}

\affil[2]{\href{mailto:mdeleon@icmat.es}{mdeleon@icmat.es}}

\affil[-]{Universidad de Alcal\'a (UAH), Departamento de F\'isica y Matem\'aticas.
Av. de Le\'on, 4A, 28805 Alcalá de Henares, Madrid, Spain} 

\affil[*]{Instituto de Ciencias Matem\'aticas (CSIC-UAM-UC3M-UCM),
C\textbackslash Nicol\'as Cabrera, 13-15, Campus Cantoblanco, UAM
28049 Madrid, Spain} 

\affil[+]{Real Academia de Ciencias Exactas, Fisicas y Naturales, C/de Valverde
22, 28004 Madrid, Spain}

\date{\today}

\begin{document}

\sloppy

\maketitle



\begin{abstract}

The aim of this paper is to study the evolution of a material point of a body by itself, and not the body as a whole. To do this, we construct a groupoid encoding all the intrinsic properties of the material point and its characteristic foliations, which permits us to define the evolution equation. We also discuss phenomena like remodeling and aging.

\end{abstract}

\maketitle

\tableofcontents

\section{Introduction}

Our approach to the theory of elasticity is based on the theory developed by Walter Noll \cite{WNOLLTHE}, in the sense that the mechanical response of a body to deformation is determined by a constitutive law, i.e., a functional that depends on the point and the derivatives of the deformation at that point. We will not consider more general situations, such as the dependence of higher order derivatives or media with microstructure. We will always remain in the scenario of what we will call simple elastic materials.

The existence of a constitutive law makes it possible to compare two points of the material using the concept of material isomorphism. That is, given two particles $X$ and $Y$ of a body $\mathcal{B}$, we will say that they are made of the same material if there exists a linear isomorphism $P_{X,Y} : T_{X} \mathcal{B} \longrightarrow T_{Y} \mathcal{B}$ such that for any other linear isomorphism $F_{Y,Z} : T_{Y} \mathcal{B} \longrightarrow T_{Z} \mathcal{B}$ the condition 
$$
W \left(X, F_{Y,Z}\cdot P_{X,Y}\right) = W \left(Y, F_{Y,Z}\right),
$$
is satisfied, where $W = W\left( X,F\right)$ is the constitutive law. In this conditions $P_{X,Y}$ is called a \textit{material isomorphism}. Here, we will use the notion of 1-jet of local diffeomorphisms which is equivalent to that of linear isomorphism, but which allows us a better mathematical treatment (see \cite{SAUND} for more details about the formalism of $1-$jets).

The collection $\Omega \left( \mathcal{B}  \right)$ of all material isomorphisms for each pair of points has an algebraic structure which is that of a groupoid; in fact, $\Omega\left( \mathcal{B}  \right)$ is a subgroupoid (named \textit{material groupoid}) of the Lie groupoid of all possible isomorphisms (not just material ones), $\Pi^1\left( \mathcal{B} , \mathcal{B}  \right)$. The material groupoid (non-differentiable in principle) makes it possible to establish conditions of uniformity and homogeneity in a very natural way. If the material groupoid is a Lie groupoid, then we can associate with it the corresponding Lie algebroid, which is the infinitesimal approximation in the same way that every Lie group has a Lie algebra associated with it. The integrability of that Lie groupoid characterizes the homogeneity of the material \cite{VMJIMM,COSVME}.

However, there are materials for which the material groupoid is not differentiable, and therefore is not a Lie groupoid. In \cite{CHARDIST,MD} we have developed a construction that generalizes the notion of Lie algebroid and yields the so-called characteristic distribution. This is a generalized distribution (in the sense of Stefan and Sussmann \cite{PS,HJS}) but involutive, thus determining generalized foliations that divide the body in a differentiable way. The characteristic distribution of a material has allowed us to introduce concepts such as graded uniformity and homogeneity, previously unknown.

In this paper, we applied a similar technique to materials that evolve with time \cite{EPSTEIN201572,MEPMDLSEG,EPSBOOK2}. In this context, we define a \textit{body-time manifold} as the fiber bundle $\mathcal{C} = \mathbb{R} \times \mathcal{B} \longrightarrow \mathbb{R}$. Furthermore, the mechanical response is assumed to be a ``\textit{differentiable curve}'' of mechanical responses, each one corresponds to the state of the body at the corresponding instant $t$. Thus, we may compare the constitutive properties of the particles at different instants of time via the so-called \textit{time-material isomorphisms}, defined analogously.

So, for a fixed particle $X$ of the body, we construct a canonically defined groupoid, called \textit{$X-$material groupoid} over $\mathbb{R}$, which consists of all time-material symmetries at the particle $X$. We construct the charateristic distributions of this groupoid, called \textit{$X-$material distribution} and \textit{$X-$body-material distribution}. By using these tools, we deals with the notions of remodeling and aging of materials. The study of these kind of processes is relevant in biological tissues, like highly stressed solid tumors \cite{tuMORS}.

The associated foliations of the material distribution permits us to prove that \textit{any process of temporal evolution of a particle may be divided in periods of remodelings in a maximal way}. These ``\textit{periods}'' corresponds, in fact, to a smooth singular foliation of $\mathbb{R}$. Furthermore, the equation characterizing the material distributions, the \textit{evolution equation}, provides a computational way to construct this foliation, characterizing remodeling and aging. The results are really interesting and we believe that they will shed light on the study of the evolution of material bodies.

%%Recently [], we have applied this technique to materials that evolve with time (or other possible parameter). In this case, we must consider the subgroupoid of the space-time isomorphisms, which are mathematically formalized as elements of the Lie groupoid of the vertical fibered space of the vector fibered space $C = \mathbb{R} \times B \longrigtarrow \mathbb{R}$. That is, we must connect not only material points but also take into account the time instants in which they are located. In a recent paper we have used the characteristic distribution construction to study remodeling and aging. We should point out that the construction of a characteristic distribution is valid in an abstract way for an arbitrary subgroupoid of a Lie groupoid, which provides us with a practically universal tool.

The paper is structured as follows. Section 2 is devoted to recall the essential elements of Noll’s constitutive theory, and it is completed with a development of a similar theory for time evolution materials in section 3. In section 4 we shall construct the material body-time groupoid, and in Section 5 we discuss remodeling and aging. An application to 
laminated liquid crystal is included in Section 6. Finally, the main ingredients on groupoids are exhibited in an Appendix, in order not to overload the previous sections of mathematical contents that can be found in some books and other papers but that we include in this way to facilitate the reading of the paper.

\section{Noll's constitutive theory}\label{partelasticmat2}

For the notion of \textit{elastic simple material} (or simply \textit{simple materials}), we will follow the theory iniciated by W. Noll \cite{WNOLLTHE}. Recommendable references for a detailed study of this topic are \cite{EPSBOOK2,JEMARS,CCWAN}.
\begin{definition}
\rm
An \textit{elastic simple material} is given by an oriented manifold $\mathcal{B}$ of dimension $3$ which may be embedded in $\mathbb{R}^{3}$. Points of $\mathcal{B}$ are called \textit{material points} or \textit{material particles} and will be denoted by capital letters ($X,Y,Z \in \mathcal{B}$)
\end{definition}
\noindent{Any open subset $\mathcal{U}$ of the manifold $\mathcal{B}$ is called a \textit{sub-body}.} A \textit{configuration} is an embedding $\phi : \mathcal{B} \rightarrow \mathbb{R}^{3}$. An \textit{infinitesimal configuration at a particle $X$} is given by the $1-$jet $j_{X,\phi \left(X\right)}^{1} \phi$ where $\phi$ is a configuration of $\mathcal{B}$. To study in detail the formalism of $1-$jets see \cite{SAUND}. Along the paper, a configuration, called \textit{reference configuration}, $\phi_{0}$ will be fixed. The open set $\mathcal{B}_{0} = \phi_{0} \left( \mathcal{B} \right)$ will be called \textit{reference state}. The local coordinates in the reference configuration will be denoted by $X^{I}$ and any other coordinates will be denoted by $x^{i}$.\\
A \textit{deformation} of the body $\mathcal{B}$ is defined as the change of configurations $\kappa = \phi_{1} \circ \phi_{0}^{-1}$ or, equivalently a diffeomorphism from the reference state $\mathcal{B}_{0}$ to any other open subset $\mathcal{B}_{1} = \phi_{1}\left( \mathcal{B} \right)$ of $\mathbb{R}^{3}$. Analogously, an \textit{infinitesimal deformation at $\phi_{0}\left(X\right)$} is given by a $1-$jet $j_{\phi_{0}\left(X\right) , \phi \left(X\right)}^{1} \kappa$ where $\kappa$ is a deformation.\\
Broadly speaking, the configuration is a way of manifesting the body into the ``\textit{real world}''. The points on the euclidean space $\mathbb{R}^{3}$ will be called \textit{spatial points} and will be denoted by lower case letters ($x,y,z\in \mathbb{R}^{3}$).\\
Following the theory developed by W. Noll \cite{WNOLLTHE}, the internal properties of the body are characterized for the so-called \textit{constitutive equations}. For \textit{elastic simple bodies}, we will assume that the constitutive law depends on a material point only on the infinitesimal deformation at that point.
\begin{definition}
\rm
The \textit{mechanical response} of a (elastic) simple material $\mathcal{B}$, in a fixed reference configuration $\phi_{0}$, is formalized as a differentiable map $W$ from the set $\mathcal{B} \times Gl \left( 3 , \mathbb{R} \right)$, where $Gl \left( 3 , \mathbb{R} \right)$ is the general linear group of $3 \times 3$-regular matrices, to a fixed (finite dimensional) vector space $V$.
\end{definition}

In this paper, we will not be interested in particular cases of the vector space $V$. However, it is relevant to specify that, in general, $V$ will be the space of \textit{stress tensors}. More particularly, the contact forces at a particle $X$ (in a fixed configuration $\phi$) are determined by a symmetric second-order tensor 
$$T_{X,\phi}: \mathbb{R}^{3} \rightarrow \mathbb{R}^{3}$$
on $\mathbb{R}^{3}$ called the \textit{stress tensor}. Then, the mechanical response is given as follows:
$$W \left( X , F \right) = T_{X,\phi},$$
where $F$ is the $1-$jet $j_{\phi_{0} \left( X \right),\phi\left( X\right)}^{1}\left( \phi \circ \phi_{0}^{-1}\right)$ at $\phi_{0} \left( X \right)$ of $\phi \circ \phi_{0}^{-1}$.\\
We should now introduce the \textit{rule of change of reference configuration}. In particular, consider  another configuration $\phi_{1}$ and the associated mechanical response $W_{1}$. Then, we will impose that,
\begin{equation}\label{1.5}
 W_{1} \left( X , F \right) = W \left( X , F \cdot C_{01} \right),
\end{equation}
for all regular matrix $F$, where $C_{01}$ is the associated matrix to the $1-$jet at $\phi_{0} \left( X \right)$ of $\phi_{1} \circ \phi_{0}^{-1}$. 
This fact is equivalent to the identity,
\begin{equation}\label{1.4}
 W \left( X , F_{0} \right) = W_{1} \left( X , F_{1} \right),
\end{equation}
for any configuration $\phi$, where $F_{i}$, $i=0,1$, is the associated matrix to the $1-$jet at $\phi_{i} \left( X \right)$ of $\phi \circ \phi_{i}^{-1}$ \cite{VMMDME}.\\

There are several equivalent ways of presenting the mechanical response. On the one hand, we may define define $W$ on the space of (local) configurations in such a way that for each configuration $\phi$ we have that
$$ W \left( j^{1}_{X, x} \phi \right) = W \left( X , F \right),$$
where $F$ is the associated matrix to the $1-$jet at $\phi_{0} \left( X \right)$ of $\phi \circ \phi_{0}^{-1}$. So, Eq. (\ref{1.4}) implies that this map \textbf{does not depend on the chosen reference configuration}.\\
On the other hand, consider $\Pi^{1}\left( \mathcal{B}, \mathcal{B}\right)$ the manifold of the $1-$jets of (local) diffeomorphisms from $\mathcal{B}$ to $\mathcal{B}$ (\cite{KMG}). Then, $W$ may be described as a differentiable map $W : \Pi^{1}\left( \mathcal{B} , \mathcal{B} \right) \rightarrow V$ from $\Pi^{1}\left( \mathcal{B} , \mathcal{B} \right)$ to the vector space $ V$ by the following identity,
\begin{equation}\label{4.1}
 W \left( j_{X,Y}^{1} \kappa\right) = W \left( X , F\right),
\end{equation}
where $F$ is the associated matrix to the $1-$jet at $\phi_{0} \left( X \right)$ of $\phi_{0}\circ \phi \circ \phi^{-1}_{0}$. If there is not danger of confusion, we will use in this paper these three ways of describing the mechanical response, indistinctly.\\
Observe that, restricting the mechanical response, any sub-body inherits the structure of elastic simple body from the body $\mathcal{B}$. A fundamental notion in the theory of W. Noll is the concept of \textit{material isomorphism}, which permits to compare the material properties of two different points.
\begin{definition}\label{1.33}
\rm
Let $\mathcal{B}$ be a body. Two material particles $X$ and $Y$ are materially isomorphic if, and only if, there exists a local diffeomorphism $\psi$ from an open subset $\mathcal{U} \subseteq \mathcal{B}$ of $X$ to an open subset $\mathcal{V} \subseteq \mathcal{B}$ of $Y$ such that $\psi \left(X\right) =Y$ and
\begin{equation}\label{4.2}
W \left( j^{1}_{Y, \kappa \left(Y\right)} \kappa \cdot j^{1}_{X,Y} \psi \right) = W \left( j^{1}_{Y, \kappa \left(Y\right)} \kappa\right),
\end{equation}
for all $j^{1}_{Y , \kappa \left(Y\right)} \kappa \in \Pi^{1}\left( \mathcal{B} , \mathcal{B} \right)$. In these conditions, $j^{1}_{X,Y} \psi$ will be called a material \textit{isomorphism from $X$ to $Y$.} A material isomorphism from $X$ to itself is called a \textit{material symmetry}. In cases where it causes no confusion we often refer to associated matrix $P$ as the material isomorphism (or symmetry).
\end{definition}

\begin{remark}
\rm
\noindent{We should notice that the elements of $\Pi^{1}\left( \mathcal{B} , \mathcal{B} \right)$ may be interpreted as linear isomorphisms $L_{X,Y}: T_{X}\mathcal{B}  \rightarrow T_{Y}\mathcal{B}$ beetwen the tangent spaces of the body $\mathcal{B}$ at two different particles $X$ and $Y$.}
\end{remark}
So, from a physical point of view, two points are materially isomorphic if their intrinsic properties are the same, i.e., they are part of the same material. In fact, we have that,
\begin{proposition}[\cite{VMMDME}]
Let $\mathcal{B}$ be a body. Two body points $X$ and $Y$ are materially isomorphic if, and only if, there exist two (local) configurations $\phi_{1}$ and $\phi_{2}$ such that
$$ W_{1} \left( X , F \right)= W_{2} \left( Y , F \right), \ \forall F,$$
where $W_{i}$ is the mechanical response associated to $\phi_{i}$ for $i=1,2$.
\end{proposition}

The set of all material symmetries at particle $X$ will be denoted by $G \left( X \right)$. So, all the symmetry groups of materially isomorphic particles are conjugated. In other words, if $X$ and $Y$ are material isomorphic we have that
$$ G \left( Y \right) = P \cdot G \left( X \right) \cdot P^{-1},$$
where $P$ is a material isomorphism from $X$ to $Y$.
\begin{definition}\label{1.17}
\rm
A body $\mathcal{B}$ is said to be \textit{uniform} if all of its body points are materially isomorphic.
\end{definition}
\noindent{Thus, a body is uniform if all the points are made of the same material.}\\\\
%%Let us consider a uniform body $\mathcal{B}$ and a fixed body point $X_{0}$. Then, for any other particle $Y$ we may find a material isomorphism from $X_{0}$ to $Y$, say $P \left( Y \right) $. Therefore, we shall construct a map 
%%$$\mathcal{P} : \mathcal{B} \rightarrow   \Pi^{1} \left( \mathbb{B} , \mathcal{B} \right)$$ 
%%(or equivalently a map $P : \mathcal{B} \rightarrow   Gl \left( 3, \mathbb{R} \right)$) consisting of material isomorphisms. Nevertheless, $P$ does not have to be a differentiable map.
%%\begin{definition}\label{1.7}
%%\rm
%%A body $\mathcal{B}$ is said to be \textit{smoothly uniform} if for each point $X \in \mathcal{B}$ there is a neighbourhood $\mathcal{U} $ around $X$ and a smooth map $P : \mathcal{U} \rightarrow Gl \left( 3, \mathbb{R} \right)$ such that for all $Y \in \mathcal{U}$ it satisfies that $P \left(Y\right)$ is a material isomorphism from $X$ to $Y$. The map $P$ is called a \textit{right (local) smooth field of material isomorphisms}. A \textit{left (local) smooth field of material isomorphisms} is defined analogously.
%%\end{definition}

In our geometrical framework, we have studied the intrinsic properties of the body without permitting changes in time. The constitutive properties of the body are frozen, and remain without any change during the whole process of deformation. Nevertheless, in several practical application the material tend to be in motion or evolve in time. A relevant example is given by the remodeling of biological tissues, like muscles or bones. We will deal with the material evolution following the model studied in the following references \cite{EPSTEIN201572,MEPMDLSEG,EPSBOOK2}.\\\\

\section{A constitutive theory for time evolution materials}

\noindent{A \textit{body-time manifold} will be given by the fibre bundle $ \mathcal{C} = \mathbb{R} \times \mathcal{B}$ over $\mathbb{R}$. We assume that time and space are absolute by simplicity, but it is not difficult to extend the theory to a general case.}
\begin{definition}
\rm
A \textit{history} is given by a fibre bundle embedding $\Phi : \mathcal{C} \rightarrow \mathbb{R} \times \mathbb{R}^{3}$ over the identity.
\end{definition}
In other words, a history $\Phi$ is characterized by a differentiable family of configurations $\phi_{t} : \mathcal{B} \rightarrow \mathbb{R}^{3}$ of $\mathcal{B}$ such that
\begin{equation}\label{eqution345}
    \phi_{t} \left( X \right) = \Phi \left( t , X \right), \ \forall t \in \mathbb{R}, \ \forall X \in \mathcal{B}.
    \end{equation}
Thus, $\Phi$ present, at each time $t$, a (possibly) different configuration $\phi_{t}$ of the body $\mathcal{B}$. Hence, at any time $t$, we may consider the $1-$jet $j_{X, \phi_{t}\left(X \right)}^{1} \phi_{t}$ which is the infinitesimal configuration at time $t$.\\
In the case of simple bodies, we assumed that for a fixed reference configuration $\phi_{0}$, the constitutive law at each body point $X$ and at each instant $t$ may be determined by one (or more) functions depending on the infinitesimal configurations $j_{X, \phi_{t}\left(X \right)}^{1} \phi_{t}$ at particle $X$ and time $t$. Therefore, it turns out to be natural to define the \textit{mechanical response} of the temporal evolution as a differentiable map,
$$ W :  \mathcal{C} \times Gl\left(3 , \mathbb{R}\right)  \rightarrow V,$$
where $V$ is again a real vector space (as above, $V$ will be generally assumed to be the space of stress tensors).\\

Once again, we have a \textit{rule of change of reference configuration}. In particular, let us consider a different configuration $\phi_{1}$ and $W_{1}$ its associated mechanical response. Then, the rule of change of reference configuration is given by,
\begin{equation}\label{1.5.2}
 W_{1} \left(t, X , F \right) = W \left( t, X , F \cdot C_{01} \right),
\end{equation}
for all regular matrix $F$ where $C_{01}$ is the associated matrix to the $1-$jet at $\phi_{0} \left( X \right)$ of $\phi_{1} \circ \phi_{0}^{-1}$. Equivalently,
\begin{equation}\label{1.4.2}
 W \left( t,X , F_{0} \right) = W_{1} \left(t, X , F_{1} \right),
\end{equation}
where $F_{i}$, $i=0,1$, is the associated matrix to the $1-$jet at $\phi_{i} \left( X \right)$ of $\phi \circ \phi_{i}^{-1}$ with $\phi$ a configuration. Thus, the mechanical response may be equivalently defined as a differentiable family of mechanical responses for the body $\mathcal{B}$ as follows,
$$W_{t}\left( X , F \right) = W\left(t, X , F \right), $$
for all $\left(t, X , F \right) \in \mathcal{C} \times Gl\left(3 , \mathbb{R}\right)$. Thus, we may see how the mechanical response is measuring the change in the constitutive properties of the material in time.\\\\
As in the case of simple bodies, there are more several equivalent ways of presenting the mechanical response. On the one hand, Eq. (\ref{1.5.2}) permit us to define $W$ over the space of (local) histories which is \textit{independent on the chosen reference configuration}. In fact, for each history $\Phi = \phi_{t}$ we will define
\begin{equation}\label{1.4.5.5}
    W \left( t , X , \Phi \right) = W \left( j^{1}_{X,x} \phi_{t} \right),
\end{equation}
On the other hand, let us consider the vertical subbundle associated to the body-time manifold $\mathcal{C}$, $\mathcal{V}$, and the set $\Phi \left(\mathcal{V}\right)$ of all linear isomorphisms between fibres of $\mathcal{V}$ (see example \ref{8} in the appendix I). Notice that, for all $\left( t, X \right) \in \mathcal{C}$, we have that, the fibre at $\left( t, X \right)$ is given by,
$$ \mathcal{V}_{\left( t, X \right)} = \{0\} \times T_{X}\mathcal{B}.$$
An element of $\Phi \left(\mathcal{V}\right)$ may be given by a triple $\left( \left( t, X \right) , \Phi \left( t , X \right) ,  j_{X, \phi_{t}\left(X \right)}^{1} \phi_{t} \right)$, where $\Phi : \mathcal{C}\rightarrow \mathcal{C}$ is an (local) isomorphism of vector bundles over the identity at $\mathbb{R}$, and $\phi_{t}: \mathcal{B} \rightarrow \mathcal{B}$ is the (local) diffeomorphism induced by $\Phi$ at the instant $t$ as follows,
$$  \Phi \left( t , X \right) = \left( t , \phi_{t}\left( X \right) \right),    \ \forall \left( t, X \right) \in \mathcal{C}.  $$
Another, less intuitive but easier way to represent an element of $\Phi \left(\mathcal{V}\right)$ is a triple $\left(t,s , j_{X,Y}^{1} \phi\right)$ with $s,t \in \mathbb{R}$, $X \in \mathcal{B}$ and $\phi$ a local automorphism on $\mathcal{B}$ from $X$ to $Y$.\\
\noindent{By using Eq. (\ref{1.4.5.5}), we may define $W$ on the space of local histories,}
$$W : \Phi \left(\mathcal{V}\right) \rightarrow V,$$
as follows,
$$ W \left( t , s , j_{X,Y}^{1} \phi\right) = W\left( t , X , \Phi \right),$$
such that
$$ \Phi \left( s , Y \right) = \left( s , \phi_{0} \circ \phi \left( Y \right) \right), \ \forall \left( s , Y \right) \in \mathcal{C},$$
where $\phi_{0}$ is the reference configuration.
\begin{definition}
\rm
Let $\mathcal{C}$ be a body-time manifold. Two body points $X$ and $Y$  at two different instants $t$ and $s$, respectively are materially isomorphic if, and only if, there exists a local diffeomorphism $\psi$ from an open subset $\mathcal{U} \subseteq \mathcal{B}$ of $X$ to an open subset $\mathcal{V} \subseteq \mathcal{B}$ of $Y$ such that $\psi \left(X\right) =Y$ and
\begin{equation}\label{1.3.second}
W \left(t,r , j^{1}_{X,Z}\left( \phi \cdot \psi \right) \right) = W \left(s,r , j^{1}_{Y,Z}\phi \right) , 
\end{equation}
for all $\left(s,r , j^{1}_{Y,Z}\phi \right) \in \Phi \left(\mathcal{V}\right)$. In these conditions, $j^{1}_{X,Y} \psi$ will be called a \textit{time-material isomorphisms} (or \textit{material isomorphisms} if there is no danger of confusion) from $\left(t,X\right)$ to $\left(s,Y\right)$. A material isomorphism from $\left(t,X\right)$ to itself is called a \textit{time-material symmetry} or \textit{material symmetry}.
\end{definition}

\begin{proposition}
Let $\mathcal{C}$ be a body-time manifold. Two body pairs $\left(t,X\right)$ and $\left(s,Y\right)$ are materially isomorphic if, and only if, there exist two (local) configurations $\phi_{1}$ and $\phi_{2}$ such that
$$ W_{1} \left( t,X , F \right)= W_{2} \left( s,Y , F \right), \ \forall F,$$
where $W_{i}$ is the mechanical response associated to $\phi_{i}$ for $i=1,2$.
\end{proposition}
Thus, the meaning of this result is the following: \textit{two body pairs $\left(t,X\right)$ and $\left(s,Y\right)$ are materially isomorphic if, and only if, $X$ and $Y$ are made of the same material at the instants $t$ and $s$, respectively.}\\
Observe that, for all $t$, the manifold $\{t\}\times \mathcal{B}$ inherits the structure of simple body by restricting the mechanical response $W$. Thus, this simple material will be called \textit{state $t$ of the body $\mathcal{B}$} and it will be denoted by $\mathcal{B}_{t}$. As long as it invites no confusion, we will refer to the simple body $\{0\} \times \mathcal{B}$ as the material body $\mathcal{B}$.\\
On the other hand, as in the case of simple bodies, the mechanical response defines an structure of material evolution on any sub-body $\mathcal{U}$ of the body $\mathcal{B}$ by restriction.\\

\section{The material body-time groupoid}

\begin{definition}
\rm
\noindent{Let us fix two material particle $X,Y \in \mathcal{B}$. Then, we will define the \textit{$\left(X,Y\right)-$material groupoid} $\Omega_{X,Y} \left( \mathbb{R} \right)$ as the set of all material isomorphisms from the particle $X$ to the particle $Y$ varying the time variable.}
\end{definition}

\noindent{Here, the relevant case is $X =Y$. Indeed, it is easy to check that $\Omega_{X,X} \left( \mathbb{R} \right)$ is a subgroupoid of $\Phi \left( \mathcal{V}\right)$. For each material point $X$, $\Omega_{X,X} \left( \mathbb{R} \right)$ is called \textit{$X-$material groupoid} and denoted by $\Omega_{X} \left( \mathbb{R} \right)$.\\
On the other hand, $\Omega_{X} \left( \mathbb{R} \right)$ may be consider as a subgroupoid of $\left(\mathbb{R}  \times \mathbb{R}\right) \times \Pi^{1} \left( \mathcal{B} , \mathcal{B}\right)_{X}^{X}$ on $\mathbb{R}$, where we are identifying $\mathbb{R}$ with $\mathbb{R} \times \{X\}$. Furthermore, the structure of Lie groupoid of $\left(\mathbb{R}  \times \mathbb{R}\right) \times \Pi^{1} \left( \mathcal{B} , \mathcal{B}\right)_{X}^{X}$ is given by,
$$\left( s,t,j_{X,X}^{1}\phi\right) \cdot \left( r,s,j_{X,X}^{1}\psi\right) = \left( r,t,j_{X,X}^{1}\left(\phi \circ \psi \right)\right), $$
for all $\left( s,t,j_{X,X}^{1}\phi\right) , \left( r,s,j_{X,X}^{1}\psi\right) \in  \mathbb{R}\times \mathbb{R} \times\Pi^{1} \left( \mathcal{B} , \mathcal{B}\right)_{X}^{X}$.\\
We will use both interpretations of $\Omega_{X} \left( \mathbb{R} \right)$ along the paper.\\
At this point it turns out to be important to highlight that \textit{the $X-$material groupoid $\Omega_{X} \left( \mathbb{R} \right)$ is an algebraic structure, canonically defined for any material evolution $\mathcal{C}$, which encodes all the constitutive properties of the whole variation in time of the material particle $X$}. Therefore, this object will be one of the most essential pillars of this paper.\\
However, the $X-$material groupoid is not (in general) a Lie subgroupoid of $\Phi \left( \mathcal{V}\right)$. So, it is an algebraic, but not differentiable, structure associated to $\mathcal{C}$. To solve this fact, we will construct the \textit{material distributions} for evolution material.\\
}
\noindent{So, we start considering a particular class of vector field on $\mathbb{R}\times \mathbb{R} \times\Pi^{1} \left( \mathcal{B} , \mathcal{B}\right)_{X}^{X}$.}
\begin{definition}
\rm
A (local) vector field $\Theta \in \frak X_{loc} \left( \mathbb{R}\times \mathbb{R} \times\Pi^{1} \left( \mathcal{B} , \mathcal{B}\right)_{X}^{X} \right)$ on $\mathbb{R}\times \mathbb{R} \times\Pi^{1} \left( \mathcal{B} , \mathcal{B}\right)_{X}^{X}$ will be called \textit{admissible} for the couple $\left( \mathbb{R}\times \mathbb{R} \times\Pi^{1} \left( \mathcal{B} , \mathcal{B}\right)_{X}^{X} ,\Omega_{X} \left( \mathbb{R} \right) \right)$ if it satisfies that,
\begin{itemize}
\item[(i)] \textbf{$\Theta$ is tangent to the $\beta-$fibres}, i.e., 
$$ \Theta \left( g \right) \in T_{g} \beta^{-1} \left( \beta \left( g \right) \right),$$
for all $g$ in the domain of $\Theta$.
\item[(ii)] \textbf{$\Theta$ is invariant by left translations}, i.e.,
$$ \Theta \left( g \right) = T_{\epsilon \left( \alpha \left( g \right) \right) } L_{g} \left( \Theta \left( \epsilon \left( \alpha \left( g \right) \right) \right) \right),$$
for all $g $ in the domain of $\Theta$.
\item[(iii)]  $TW_{X}\left( \Theta \right) = 0$, where $W_{X}$ is the restriction of $W$ to $\mathbb{R}\times \mathbb{R} \times\Pi^{1} \left( \mathcal{B} , \mathcal{B}\right)_{X}^{X}$.
\end{itemize}
\end{definition}

In other words, the $X-$material distribution of $\mathcal{C}$ is generated by the left-invariant vector fields $\Theta$ on $\mathbb{R}\times \mathbb{R} \times\Pi^{1} \left( \mathcal{B} , \mathcal{B}\right)_{X}^{X}$ such that
$$
    T W_{X} \left( \Theta \right) = 0
$$

Let $\Theta$ be a left-invariant vector field on $\mathbb{R}\times \mathbb{R} \times \Pi^{1} \left( \mathcal{B} , \mathcal{B}\right)_{X}^{X}$. Then, locally,

\begin{equation}
    \Theta  \left(t,s, y^{i}_{j}\right)   =   \lambda\dfrac{\partial}{\partial  t } +  y^{i}_{l} \Theta^{l}_{j}\dfrac{\partial}{\partial y^{i}_{j} },
\end{equation}
respect to a local system of coordinates $\left(t,s, y^{i}_{j}\right)$ on $ \mathbb{R}\times \mathbb{R} \times \Pi^{1} \left( \mathcal{U} , \mathcal{U}\right)_{X}^{X}$ with $\mathcal{U}$ an open subset of $\mathcal{B}$ with $X \in \mathcal{U}$. Then, $\Theta$ is an admissible vector field for the couple $\left( \mathbb{R}\times \mathbb{R} \times \Pi^{1} \left( \mathcal{B} , \mathcal{B}\right)_{X}^{X}, \Omega_{X} \left( \mathbb{R} \right)\right)$ if, and only if, the following equations hold,
\begin{equation}\label{Xmatgroup323456}
    \lambda\dfrac{\partial W_{X}}{\partial  t } + y^{i}_{l} \Theta^{l}_{j}\dfrac{\partial   W_{X}}{\partial  y^{i}_{j} } = 0
\end{equation}
Observe that, here $\lambda$ and $\Theta^{i}_{j}$ are functions depending on $t$. This linear equation will be called the \textit{evolution equation of $X$}, and it will play a fundamental role in the paper.\\

Let us consider the source map of $\mathbb{R}\times \mathbb{R} \times\Pi^{1} \left( \mathcal{B} , \mathcal{B}\right)_{X}^{X}$, which is given by the first projection $\alpha: \mathbb{R}\times \mathbb{R} \times\Pi^{1} \left( \mathcal{B} , \mathcal{B}\right)_{X}^{X} \rightarrow \mathbb{R}$. On the other hand, the identity map 
$$\epsilon: \mathbb{R} \rightarrow \mathbb{R}\times \mathbb{R} \times\Pi^{1} \left( \mathcal{B} , \mathcal{B}\right)_{X}^{X}$$
satisfies that
$$\epsilon \left( t\right) = \left( t,t, j_{X,X}^{1}Id_{\mathcal{B}}\right)$$
where $Id_{\mathcal{B}}$ is the identity on $\mathcal{B}$.\\
Then, it may be constructed another differentiable distribution on $\mathbb{R}$, $A \Omega_{X} \left( \mathbb{R} \right)^{\sharp}$, characterized by the following diagram

\begin{large}
\begin{center}
 \begin{tikzcd}[column sep=huge,row sep=huge]
\mathbb{R}\times \mathbb{R} \times\Pi^{1} \left( \mathcal{B} , \mathcal{B}\right)\arrow[r, "A \Omega_{X} \left( \mathbb{R} \right)^{T}"] &\mathcal{P} \left( T \left(\mathbb{R}\times \mathbb{R} \times\Pi^{1} \left( \mathcal{B} , \mathcal{B}\right)_{X}^{X} \right)\right) \arrow[d, "T\alpha"] \\
 \mathbb{R} \arrow[u,"\epsilon"] \arrow[r,"A \Omega_{X} \left( \mathbb{R} \right)^{\sharp}"] &\mathcal{P} \left( T \mathbb{R} \right)
 \end{tikzcd}
\end{center}
\end{large}
where, for each set $C$, $\mathcal{P} \left( C \right)$ is its power set. So, it satisfy the following identities,
$$ A \Omega_{X} \left( \mathbb{R} \right)^{\sharp}_{t} = T\alpha \left( A \Omega_{X} \left( \mathbb{R} \right)^{T}_{\epsilon\left( t \right)} \right), \ \forall t \in \mathbb{R}$$
where $A \Omega_{X} \left( \mathbb{R} \right)^{\sharp}_{t} $ (resp. $A \Omega_{X} \left( \mathbb{R} \right)^{T}_{\epsilon\left( t \right)}$) is the fibre of $A \Omega_{X} \left( \mathbb{R} \right)^{\sharp}$  (resp. $A \Omega_{X} \left( \mathbb{R} \right)^{T}$) at $t$ (resp. $\epsilon\left( t \right)$).\\
 The distribution $A \Omega_{X} \left( \mathbb{R}\right)^{\sharp}$ will be called \textit{$X-$body-material distribution}. In particular, the $X-$body-material distribution is generated by the vector fields on $\mathbb{R}$,
 $$\Theta^{\sharp}\left( t \right)=\lambda\dfrac{\partial}{\partial  t }, $$
such that, there exists an admissible vector field for the couple $\left( \mathbb{R}\times \mathbb{R} \times \Pi^{1} \left( \mathcal{B} , \mathcal{B}\right)_{X}^{X}, \Omega_{X} \left( \mathbb{R} \right)\right)$ given by
$$   \Theta  \left(t,s, y^{i}_{j}\right)   =  \lambda\dfrac{\partial}{\partial  t } +  y^{i}_{l} \Theta^{l}_{j}\dfrac{\partial}{\partial y^{i}_{j} },$$
where $\Theta^{i}_{j}$ are functions depending on $t$.\\
In other words, the distribution $A \Omega_{X} \left( \mathbb{R}\right)^{\sharp}$ is generated by the scalar functions $\lambda$ in such a way that there exist nine differentible scalar functions $\Theta^{l}_{j}$ satisfying the evolution equation of $X$ (\ref{Xmatgroup323456}), i.e.,
$$\lambda\dfrac{\partial W_{X}}{\partial  t } + y^{i}_{l} \Theta^{l}_{j}\dfrac{\partial   W_{X}}{\partial  y^{i}_{j} } = 0
$$
Thus, \textit{to construct both the material distributions is reduced to solve the linear equation (\ref{Xmatgroup323456})}.\\\\
\noindent{It is important to say that both distributions are singular (their leaves may have different dimension). In fact, the material distributions of the material evolution may be seen as a particular example of the so-called \textit{characteristic distributions} \cite{VMMDME,CHARDIST}.}
\begin{theorem}\label{10.24}
Let $\mathcal{C}$ be a body-time manifold. Then, the $X-$material distribution $A \Omega_{X} \left( \mathbb{R}\right)^{T}$ and the $X-$body-material distribution $A \Omega_{X} \left( \mathbb{R}\right)^{\sharp}$ are integrable and their associated foliations $\overline{\mathcal{I}}$ and $\mathcal{I}$ are foliations of $\mathbb{R}\times \mathbb{R} \times \Pi^{1} \left( \mathcal{B} , \mathcal{B}\right)_{X}^{X}$ and $\mathbb{R}$, respectively. Furthermore, $\Omega_{X} \left( \mathbb{R}\right)$ is a union of leaves of $\overline{\mathcal{I}}$.
\end{theorem}
This result is proved as a consequence of the celebrated Stefan-Sussman's theorem \cite{PS,HJS} which deals with the integrability of singular distributions.\\
So, the distribution $A \Omega_{X} \left( \mathbb{R}\right)^{T}$ (resp. $A \Omega_{X} \left( \mathbb{R}\right)^{\sharp}$) is the tangent distribution of a smooth (possibly) singular foliation $\overline{\mathcal{I}}$ (resp. $\mathcal{I}$). Each leaf at an element $g \in \mathbb{R}\times \mathbb{R} \times \Pi^{1} \left( \mathcal{B} , \mathcal{B}\right)_{X}^{X}$ (resp. an instant $t$) is denoted by $\overline{\mathcal{I}} \left( g \right)$ (resp. $\mathcal{I} \left( t \right)$). Furthermore, the family of the leaves of $\overline{\mathcal{I}}$ (resp. $\mathcal{I}$) at elements of $\Omega_{X} \left( \mathbb{R}\right)$ (resp. at different instants) is called the \textit{$X-$material foliation} (resp. \textit{$X-$body-material foliation}).\\

\begin{theorem}\label{10.20}
For each instant $t \in \mathbb{R}$ there exists a transitive Lie subgroupoid $\Omega_{X} \left( \mathcal{I}\left(t\right)\right)$ of $\mathbb{R}\times \mathbb{R} \times \Pi^{1} \left( \mathcal{B} , \mathcal{B}\right)_{X}^{X}$ with base $\mathcal{I} \left( t \right)$.

\end{theorem}

Thus, we have divided $\mathbb{R}$ into leaves $\mathcal{I} \left( t \right)$ which have a ``\textit{maximal}'' structure of transitive Lie subgroupoids of $\mathbb{R}\times \mathbb{R} \times \Pi^{1} \left( \mathcal{B} , \mathcal{B}\right)_{X}^{X}$. In fact, we have that
\begin{equation}\label{10.17}
 \Omega_{X} \left( \mathcal{I}\left(t\right)\right)= \sqcup_{ \overline{g}\in \overline{\mathcal{I}} \left(  t,t,j_{X,X}^{1}Id_{\mathcal{B}}  \right)} \overline{\mathcal{I}} \left( s,s,j_{X,X}^{1}Id_{\mathcal{B}}\right),
\end{equation}
where $\overline{g}= \left( s,t,j_{X,X}^{1}\phi\right)$. In other words, $\Omega_{X} \left( \mathcal{I}\left(t\right)\right)$ may be depicted as a disjoint union of leaves of $\overline{\mathcal{I}}$ at the identities. Furthermore, $\Omega_{X} \left( \mathcal{I}\left(t\right)\right)$ may be equivalently defined as the smallest transitive subgroupoid of $\Omega_{X} \left( \mathbb{R}\right)$ which contains $\overline{\mathcal{I}}\left(  t,t,j_{X,X}^{1}Id_{\mathcal{B}}  \right)$. Observe that the $\beta-$fibre of this groupoid at an instant $s \in \mathcal{I} \left( t \right)$ is given by $\overline{\mathcal{I}} \left( s,s,j_{X,X}^{1}Id_{\mathcal{B}}\right)$.\\

\begin{corollary}\label{10.39}
Let $\mathcal{H}$ be a foliation of $\mathbb{R}$ such that for each $t \in \mathbb{R}$ there exists a transitive Lie subgroupoid $\Omega_{X} \left( \mathcal{H}\left(t\right)\right)$ of $\mathbb{R}\times \mathbb{R} \times \Pi^{1} \left( \mathcal{B} , \mathcal{B}\right)_{X}^{X}$ over the leaf $\mathcal{H} \left( t \right)$ contained in $\Omega_{X} \left( \mathbb{R}\right)$ whose family of $\beta-$fibres defines a foliation on $\mathbb{R}\times \mathbb{R} \times \Pi^{1} \left( \mathcal{B} , \mathcal{B}\right)_{X}^{X}$. Then, the $X-$body-material foliation $\mathcal{I}$ is coarser than $\mathcal{H}$, i.e.,
$$ \mathcal{H} \left( t \right) \subseteq \mathcal{I} \left(t \right) , \ \forall t \in \mathbb{R}.$$
Futhermore, it satisfies that
$$ \Omega_{X} \left( \mathcal{H}\left(t\right)\right) \subseteq \Omega_{X} \left( \mathcal{I}\left(t\right)\right).$$
\end{corollary}
This result gives us an idea about the \textit{maximality condition} which satisfy the $X-$material foliation and $X-$body-material foliation.

\section{Remodeling and aging}

\noindent{As a particular case of evolution, arise the so-called \textit{remodeling}. }

\begin{definition}\label{1.17.2.se}
\rm
Let $\mathcal{C}$ be a body-time manifold. A material particle $X \in \mathcal{B}$ is presenting a \textit{remodeling} when any two instants in its history are connected by a material isomorphism, , i.e., all the points at $\mathbb{R} \times \{X\}$ are connected by material isomorphisms. $\mathcal{C}$ is presenting a \textit{remodeling} when all the material points suffer a remodeling. Growth and resorption are given by a remodeling with volume increase or volume decrease of the material body $\mathcal{B}$. 
%%%The body-time manifold $\mathcal{C}$ is presenting an \textit{aging} when it is not a remodeling. A special kind of aging is given by the \textit{uniform aging}, i.e., for each $t \in \mathbb{R}$ all the points $\left( t , X \right) \in \mathcal{C}$ are isomorphic.
%$\mathcal{C}$ is a \textit{}smooth remodeling} if for each point $\left( t ,X \right) \in \mathcal{C}$ there is an infinitesimal neighbourhood $\mathcal{U} $ around $\left( t , X \right)$ such that for all $\left( s, Y \right) \in \mathcal{U}$ there exists a smooth field of material isomorphisms $\mathcal{P}$ from $\epsilon \left(t, X \right)$ to a material isomorphism $L_{ \left( s, Y\right) , \left( t,X \right)}$. {\color{norange}{\textbf{Smooth growth}}} and {\color{norange}{\textbf{smooth resorption}}} are defined in a similar way.
\end{definition}
Broadly speaking, a material particle is presenting a remodeling when the intrinsic properties does not vary in time, i.e., the material is does not suffer any kind of \textit{aging}.\\
This special case of evolution occurs in biological tissues \cite{RODRIGUEZ1994455}. Wolff's law of trabecular architecture of bones (see for instance \cite{TURNER19921}) is a relevant example. Here, trabeculae are assumed to change their orientation following the principal direction of stress. It is important to note that the fact of that the material body remains materially isomorphic with the time does not preclude the possibility of adding (growth) or removing (resorption) material, as long as the material added is \textit{of the same type}.
\begin{proposition}
Let $\mathcal{C}$ be a body-time manifold. A material particle $X \in \mathcal{B}$ is presenting a remodeling if, and only if, the $X-$material groupoid $\Omega_{X}\left(  \mathbb{R}\right)$ is transitive. $\mathcal{C}$ is presenting a remodeling if, and only if, for all material point $X$, the $X-$material groupoid $\Omega_{X}\left(  \mathbb{R}\right)$ is transitive.
\end{proposition}

Observe that, an evolution could happens to be a remodeling even in the case of non-uniform materials, for example, for laminate of granular materials (\cite{EPST}).\\

\noindent{On the other hand, the definition of remodeling is pointwise. In other words, consider a material particle $X_{0}$ which presents a remodeling. Then, there exists a map,}
\begin{equation}\label{1.564.34.5.second}
    \mathcal{P}: \mathbb{R} \rightarrow \Omega_{X_{0}}\left( \mathbb{R}\right)_{t_{0}}
\end{equation}
such that, for all $t \in \mathbb{R}$, $\mathcal{P} \left(t\right)$ is a material isomorphism from $\left( t_{0} , X_{0}\right)$ to $\left( t , X_{0}\right)$ for a fixed time $t_{0}$. These kind of maps are called \textit{right local (smooth) remodeling process at $t$}. We may define analogously the \textit{leftt local (smooth) remodeling process at $t$}.\\
The important fact here is that, \textit{the existence of these kind of maps does not guaratee its differentiability}.
\begin{definition}\label{1.7.2.second}
\rm
Let be a body-time manifold $\mathcal{C}$. A material point $X_{0}$ is said to be presenting a \textit{smooth remodeling} if for each point $t \in \mathbb{R}$ there is an interval $I $ around $t$ and a smooth map $P : I \rightarrow Gl \left( 3, \mathbb{R} \right)$ such that for all $s \in I$ it satisfies that $P \left(s\right)$ is a material isomorphism from $\left(t,X_{0} \right)$ to $\left(s,X_{0} \right)$. The map $P$ is called a \textit{right (local) smooth remodeling process at} $X_{0}$. A \textit{left (local) smooth remodeling process at} $X_{0}$ is defined in a similar way.
\end{definition}

Equivalently, $X_{0}$ is presenting a smooth remodeling if, and only if, for any two instants $t$ and $s$, there are two open intervals $I$ and $J$ of $t$ and $s$ respectively and a differentiable map
$$\mathcal{P} : I \times J  \rightarrow \Omega_{X_{0}} \left( \mathbb{R} \right) \subseteq \Phi \left( \mathcal{V}  \right),$$
which is a  section of the anchor map $\left( \alpha , \beta \right): \Phi \left( \mathcal{V}  \right) \rightarrow \mathcal{C}\times \mathcal{C} $ of $\Phi \left( \mathcal{V}  \right)$. When $t=s$ we may assume $I =J$ and $\mathcal{P}$ is a\textit{ morphism of groupoids over the identity map}, i.e.,

$$ \mathcal{P} \left( z,t \right) = \mathcal{P} \left(  r , t\right) \mathcal{P} \left( z, r \right), \ \forall t,r,z \in I.$$
These kind of maps are called \textit{local (smooth) remodeling processes.}\\

\noindent{Since the mechanical response $W$ is a continuous map, for all material point $X \in \mathcal{B}$ and all instant $t$, the symmetry group $ \Omega_{X}\left(   \mathbb{R} \right)_{t}^{t}$ is a Lie subgroup of $\Phi \left( \mathcal{V} \right)_{\left(t,X\right)}^{\left( t,X\right)}$.}

\begin{proposition}\label{4.4.second2324}
Let be a body-time manifold $\mathcal{C}$ and a material point $X_{0}$. $X_{0}$ is presenting a smooth remodeling if, and only if, $\Omega_{X_{0}} \left( \mathbb{R}\right)$ is a transitive Lie subgroupoid of $\Phi \left( \mathcal{V} \right)$.
\begin{proof}
Suppose that $X_{0}$ is presenting a smooth remodeling. Let be a triple $\left( s,t , j_{X_{0},X_{0}}^{1} \phi \right) \in \Omega_{X_{0}} \left( \mathbb{R} \right)$ and a local (smooth) remodeling process through $\left( s,t , j_{X_{0},X_{0}}^{1} \phi \right)$,
$$\mathcal{P} : I \times J  \rightarrow \Omega_{X_{0}} \left( \mathbb{R} \right) \subseteq \Phi \left( \mathcal{V}  \right),$$
with $s \in I$ and $t \in J$. Then, we will construct the following one-to-one map,

$$
\begin{array}{rccl}
\Psi_{I,J} : & \Omega_{X_{0}} \left( I,J\right) & \rightarrow & \mathbb{R} \times \mathbb{R} \times \Omega \left( \mathcal{C} \right)_{\left(s,X_{0}\right)}^{\left( t,X_{0}\right)}\\
&\left( k,l , j_{X_{0},X_{0}}^{1} \psi \right) &\mapsto &  \left( k, l  , \mathcal{P} \left( l , t \right) \left[ \left( k,l , j_{X_{0},X_{0}}^{1} \psi \right) \right] \mathcal{P} \left( s, k \right)\right)
\end{array}
$$

\noindent{where $\Omega_{X_{0}} \left( I,J\right)$ is the set of material isomorphisms at $X_{0}$ from instants at $I$ to instants at $J$. Taking into account that $\Omega \left( \mathcal{C} \right)_{\left(s,X_{0}\right)}^{\left( t,X_{0}\right)}$ is a differentiable manifold. Thus, we can endow $\Omega_{X_{0}} \left( \mathbb{R} \right)$ with a differentiable structure of a manifold. Finally, a proof of the converse statement may be found in \cite{KMG}).}
\end{proof}
\end{proposition}
Next, it is crucial to understand the importance of this result. In fact, it expresses how the property of \textit{smoothness} on the remodeling processes affects to the set of material isomorphisms. Therefore, \textit{the study of smooth remodeling is reduced to study the differentiability of the set of material isomorphisms and it is not necessary to find specific smooth remodeling processes which, in general, is obviously much more difficult.}\\
Notice that the existence of remodeling processes is not canonical. In fact, for a (local) remodeling process at a particle $X_{0}$
$$\mathcal{P} : I \times J  \rightarrow \Omega_{X_{0}} \left( \mathbb{R} \right) \subseteq \Phi \left( \mathcal{V}  \right),$$
any other remodeling process $\mathcal{Q}$ satisfies that
$$\mathcal{Q} \left(t,s \right) \in   \mathcal{P} \left( t_{0},s \right) \cdot  \Omega \left( \mathcal{C} \right)_{\left(t_{0},X_{0}\right)}^{\left( t_{0},X_{0}\right)} \cdot \mathcal{P} \left( t,t_{0} \right)$$
Thus, the symmetry groups provide a measure of the degree of freedom available in the choice of the remodeling process.\\\\

\noindent{Let $X$ be a particle at $\mathcal{B}$. Consider the $X-$material distribution $A \Omega_{X} \left(\mathbb{R} \right)^{T}$ with its associated $X-$body-material distribution $A \Omega_{X} \left( \mathbb{R}\right)^{\sharp}$. Let us also consider the $X-$material foliation and the $X-$body-material foliation $\overline{\mathcal{I}}_{X}$ and $\mathcal{I}_{X}$, respectively.}\\
Notice that, the foliation $\mathcal{I}_{X}$ is a  foliation of $\mathbb{R}$. Hence, each of the leaves are open intervals (dimension $1$) or single instants (dimension $0$).\\
Taking into account Theorem \ref{10.20} for each instant $t$, there exists a transitive Lie subgroupoids $\Omega_{X} \left( \mathcal{I}_{X}\left(  t \right) \right)$ of $\Phi \left( \mathcal{V} \right)$ over $\mathcal{I}_{X}\left(  t \right)$.

\begin{theorem}\label{14.1.second323.evolution}
Let be a body-time manifold $\mathcal{C}$ and a material point $X$.
The $X-$body-material foliation $\mathcal{I}_{X}$ divides $\mathbb{R}$ into maximal smooth remodeling processes.
\end{theorem}
So, this theorem shows the intuitive assertion of that, even in the case of a process of \textit{aging} (changes in the intrinsic properties, Definition \ref{1.17.2}), the temporal evolution of a material body $\mathcal{B}$ may be divided into a ``\textit{maximal}'' separation of different period of remodeling (no changes in the intrinsic properties) processes.\\
We should notice that, in this case, ``maximal'' means that any other foliation $\mathcal{H}$ of $\mathbb{R}$ by smooth remodeling processes is thinner than $\mathcal{I}_{X}$, i.e.,
$$ \mathcal{H} \left( t \right) \subseteq  \mathcal{I}_{X} \left( t \right)   , \ \forall t \in  \mathbb{R}.$$
\begin{corollary}\label{anothercorollary23}
Let be a body-time manifold $\mathcal{C}$ and a material point $X$. $X$ presents a smooth remodeling process if, and only if, $dim  \left( A \Omega_{X} \left( \mathbb{R}\right)^{\sharp}_{t}\right) = 1 $ for all instant $t$, with $A \Omega_{X} \left( \mathbb{R}\right)^{\sharp}_{t}$ the fibre of $A \Omega_{X} \left( \mathbb{R}\right)^{\sharp}$ at $t$.
\end{corollary}

Hence, to evaluate if the a material particle presents a smooth remodeling process we have to study the solutions of the evolution equation (\ref{Xmatgroup323456}),
\begin{equation}
    \lambda\dfrac{\partial W_{X}}{\partial  t } + y^{i}_{l}\Theta^{l}_{j}\dfrac{\partial   W_{X}}{\partial  y^{i}_{j} } = 0
\end{equation}
In particular, if we were able to find a solution of Eq. (\ref{Xmatgroup323456}) with $\lambda \neq 0$, the evolution would present a smooth remodeling process.\\\\

These last two results could be consider the core results of the theory of this paper. The first one claims that that, \textbf{for any evolution}, there exists a maximal partition (foliation) of the time given by remodelings. The other result give a ``\textit{computable path}'' of calculating the foliations (solving the evolution equation). Several interesting examples of remodeling processes may be found in the literature. In particular, in \cite{EPSBOOK2} it is used a model for orthotropic solids in which the tensor $P$ is proper orthogonal at all times. This model simulates an evolution law in trabeculae bones.

%%\begin{corollary}
%%Let be a body-time manifold $\mathcal{C}$. $\mathcal{C}$ presents a remodeling process if, and only if, $dim  \left( A \Omega_{X} \left( \mathbb{R}\right)^{B}_{t}\right) = 1 $ for all instant $t$ and particle $X$, with $A \Omega_{X} \left( \mathbb{R}\right)^{B}_{t}$ the fibre of $A \Omega_{X} \left( \mathbb{R}\right)^{B}$ at $t$.
%%\end{corollary}

\begin{definition}\label{1.17.2}
\rm
Let $\mathcal{C}$ be a body-time manifold. A material particle $X \in \mathcal{B}$ is presenting a \textit{aging} when it is not presenting a remodeling, i.e., not all the instants are connected by a material isomorphism. $\mathcal{C}$ is a \textit{process of aging} if it is not a process of remodeling.
%%%The body-time manifold $\mathcal{C}$ is presenting an \textit{aging} when it is not a remodeling. A special kind of aging is given by the \textit{uniform aging}, i.e., for each $t \in \mathbb{R}$ all the points $\left( t , X \right) \in \mathcal{C}$ are isomorphic.
%$\mathcal{C}$ is a \textit{}smooth remodeling} if for each point $\left( t ,X \right) \in \mathcal{C}$ there is an infinitesimal neighbourhood $\mathcal{U} $ around $\left( t , X \right)$ such that for all $\left( s, Y \right) \in \mathcal{U}$ there exists a smooth field of material isomorphisms $\mathcal{P}$ from $\epsilon \left(t, X \right)$ to a material isomorphism $L_{ \left( s, Y\right) , \left( t,X \right)}$. {\color{norange}{\textbf{Smooth growth}}} and {\color{norange}{\textbf{smooth resorption}}} are defined in a similar way.
\end{definition}
Clearly, if the material response is not preserved along the time via material isomorphism, the constitutive properties are changing with the time. Of course, like in the case of remodeling, it is reasonable the existence of \textit{smooth aging}. However, since we do not have something like a map representing the ``\textit{process of aging}", it is not easy to glimpse how to present a mathematical definition of this phenomenon. The use of the material groupoids will solve this problem.\\

\begin{proposition}\label{auxprop4342}
Let $\mathcal{C}$ be a body-time manifold. A material particle $X \in \mathcal{B}$ is presenting an aging if, and only if, the $X-$material groupoid $\Omega_{X}\left(  \mathbb{R}\right)$ is not transitive. $\mathcal{C}$ is presenting an aging if, and only if, for some material point $X$, the $X-$material groupoid $\Omega_{X}\left(  \mathbb{R}\right)$ is not transitive.
\end{proposition}
\noindent{Then, we are ready to present a proper definition of \textit{smooth aging}.}

\begin{definition}
Let $\mathcal{C}$ be a body-time manifold. A material particle $X \in \mathcal{B}$ is presenting a \textit{smooth aging} if the $X-$material groupoid $\Omega_{X}\left(  \mathbb{R}\right)$ is a Lie subgroupoid of $\Phi \left( \mathcal{V} \right)$ which is not transitive.
\end{definition}

In this way, by using the tool of groupoids, we are able to present a coherent definition of smooth aging.\\

\noindent{Let us fix a material particle $X \in \mathcal{B}$. Consider the $X-$material distribution  $A \Omega_{X} \left(\mathbb{R} \right)^{T}$. Remember $A \Omega_{X} \left(\mathbb{R} \right)^{T}$ is generated by the left-invariant vector fields $\Theta$ on $\mathbb{R}\times \mathbb{R} \times\Pi^{1} \left( \mathcal{B} , \mathcal{B}\right)_{X}^{X}$ such that}
$$
    T W_{X} \left( \Theta \right) = 0
$$
Let be the $X-$body-material distribution $A \Omega_{X} \left( \mathbb{R}\right)^{\sharp}$ and the associated foliations: the $X-$material foliation $\overline{\mathcal{I}}_{X}$ and $X-$body-material foliation $\mathcal{I}_{X}$.\\
By using Theorem \ref{14.1.second323.evolution}, we may in fact proves the following result.
\begin{proposition}
Let be a body-time manifold $\mathcal{C}$ and a material point $X$. $X$ presents a smooth aging process if, and only if, the dimension of the fibres of $A \Omega_{X} \left( \mathbb{R}\right)^{T}$ is constant and 
\begin{equation}\label{anopthermorequat2343}
dim  \left( A \Omega_{X} \left( \mathbb{R}\right)^{\sharp}_{t} \right)  =0, 
\end{equation}
for all instants $t$, with $A \Omega_{X} \left( \mathbb{R}\right)^{\sharp}_{ t }$ the fibre of $A \Omega_{X} \left( \mathbb{R}\right)^{\sharp}$ at $t$.
\begin{proof}
If $\Omega_{X} \left( \mathbb{R}\right)$ is a Lie groupoid, the dimension of the leaves of $A \Omega_{X} \left( \mathbb{R}\right)^{\sharp}$ and $A \Omega_{X} \left( \mathbb{R}\right)^{T}$ is constant. However, the dimension $A \Omega_{X} \left( \mathbb{R}\right)^{T}$ cannot be $1$, because it would be transitive. Hence, it has to happens that
$$
dim  \left( A \Omega_{X} \left( \mathbb{R}\right)^{\sharp}_{t} \right)  =0, 
$$
for all instants $t$.
\end{proof}
\end{proposition}
So, as a consequence, we have proved that there is only one possibility of obtaining a smooth aging: \textit{the maximal partition of the time into remodeling periods does not contain any interval}. Furthermore, with this result, again, we have a computational way to check if a material particle presents present a \textit{smooth aging}.\\
Therefore, as a summary we have a linear equation, the evolution equation (\ref{Xmatgroup323456}),
$$
    \lambda\dfrac{\partial W_{X}}{\partial  t } + y^{i}_{l}\Theta^{l}_{j}\dfrac{\partial   W_{X}}{\partial  y^{i}_{j} } = 0
$$
including ten variables whose solutions divide the evolution in smooth remodeling and smooth aging. Hence, the implementation of the material groupoids and their associated distributions has permitted us to prove the existence of a \textit{unique} and \textit{canonically defined} ``\textit{partition}'' of the time into smooth remodeling processes . On the other hand, it has generated a linear equation, the evolution equation (\ref{Xmatgroup323456}), whose solutions permit us to construct this partition.

\section{An application: Laminated liquid crystal}

We will consider a modified model of the so-called \textit{laminated liquid crystals} \cite{VMMDME,COLE,CCWANTHIRD}. The structure will be given by a body $\mathcal{B}$ together with a reference configuration $\psi_{0}$, where $\mathcal{B}$ is the open ball $\mathcal{B}_{0} = B_{r}\left( 0 \right)$ in $\mathbb{R}^{3}$ of radius $r$ and center $0 \in \mathbb{R}^{3}$. Furthermore, $\psi_{0}$ induces on $\mathcal{C}_{0} = \mathbb{R}\times \mathcal{B}_{0}$ a mechanical response $\mathcal{W}$ determined by the following objects:
\begin{itemize}
\item[\textbf{(i)}]$\ $  A fixed vector field $e$ on $\mathcal{B}_{0}$ such that $e \left( X \right) \neq 0$ for all $X \in \mathcal{B}_{0}$. \\

\item[\textbf{(ii)}]$\ $ Two differentiable maps $r,J: \Pi^{1}\left( \mathcal{B}_{0} , \mathcal{B}_{0} \right) \rightarrow \mathbb{R}$ in the following way
\begin{itemize}
\item $r \left( j^{1}_{X,Y} \phi \right) = g \left( Y \right) \left( T_{X} \phi \left(  e \left( X  \right)\right),  T_{X} \phi \left( e \left( X  \right)\right) \right) + \parallel X \parallel^{2}$
\item $ J \left( j^{1}_{X,Y} \phi \right) = det \left( F \right)$
\end{itemize}
\noindent
where $F$ is the Jacobian matrix of $ \phi$ with respect to the canonical basis of $\mathbb{R}^{3}$ at $X$, $g$ is a Riemannian metric on $\mathcal{B}_{0}$ and $\parallel \cdot \parallel$ the Euclidean norm of $\mathbb{R}^{3}$.\\

\item[\textbf{(iii)}]$\ $ A differentiable immersion $\widehat{W}: \mathbb{R}^{2} \rightarrow V$, with $V$ a finite-dimensional $\mathbb{R}-$vector space.
\item[\textbf{(iv)}]$\ $ A function $\mu$ depending on time, which determines the material stiffness coefficient.
\end{itemize}

Thus, these objects induce a structure of material evolution on $\mathcal{B}_{0}$ by considering the mechanical response $ \mathcal{W} : \Phi \left( \mathcal{V}_{0} \right) \rightarrow V$, where $\mathcal{V}_{0}$ is the vertical bundle of $\mathcal{B}_{0}$, as the composition  
$$\mathcal{W} \left( t,s,j^{1}_{X,Y} \phi \right) = \mu \left( t \right) \cdot \left[  \widehat{W} \circ  \left( r  , J \right) \left( j^{1}_{X,Y} \phi \right) \right].$$
We are assuming that the density remains constant. The physical meaning of this assumption is that the aging of the body is uniquely determined by its qualitative degradation in terms of the components (like magnesium in bone) present in very small quantities. By using the results in \cite{VMMDME}, it is easy to prove that in these conditions, except at the origin, all the states of the material $\mathcal{B}_{0}$ are divided in concentric spheres of (possibly) different simple liquid crystal. However, the structure of liquid crystal may be driven by a process of aging from one instant to another.\\

Let us now fix the canonical (global) coordinates $\left(t, X^{I}\right)$ of $\mathbb{R}\times \mathbb{R}^{3}$. Then, these coordinates induce a (canonic) identification $T \left(\mathbb{R}\times \mathcal{B}_{0} \right) \cong  \left[ \mathbb{R}\times\mathcal{B}_{0} \right] \times \mathbb{R}\times \mathbb{R}^{3}$. By using this identification any vector $\left( v_{t} , V_{X}\right) \in T_{t,X}\left(\mathbb{R}\times \mathcal{B}_{0} \right)$ can be equivalently expressed as $ \left(t, X ,v, V^{i} \right)$ in $\left[ \mathbb{R}\times\mathcal{B}_{0} \right] \times \mathbb{R}\times \mathbb{R}^{3}$. For the same reason, $r$ can be written as follows:
$$r \left( j^{1}_{X,Y} \phi \right) = g \left( Y \right) \left( F^{j}_{L} e^{L} \left( X  \right),  F^{j}_{L}e^{L} \left(X\right) \right) +  \parallel X \parallel^{2},$$
where $e \left( X \right) = \left( X, e^{I} \left( X \right) \right) \in \mathcal{B}_{0} \times \mathbb{R}^{3}$. Both expressions will be used with the same notation as long as there is no danger of confusion.\\

Now, we want to study the evolution equation (\ref{Xmatgroup323456}),
$$\lambda\dfrac{\partial W_{X}}{\partial  t } + y^{i}_{l} \Theta^{l}_{j}\dfrac{\partial   W_{X}}{\partial  y^{i}_{j} } = 0
$$
Notice that, for each $U= \left(U^{j}_{i}\right) \in gl \left( 3, \mathbb{R} \right)$ and $v=\left(v^{i}\right) \in \mathbb{R}^{3}$ we have that,
\begin{eqnarray*}
&\textbf{(i)}&\ \ \  \dfrac{\partial r}{\partial F_{\vert j_{X,X}^{1}\phi}} \left(  U  \right) = 2   \ g \left( X\right)\left(F^{j}_{L} e^{L} \left( X  \right) ,U^{j}_{L} e^{L} \left( X  \right) \right)\\
&\textbf{(ii)}&\ \ \  \dfrac{\partial J}{\partial F_{\vert j_{X,X}^{1}\phi}} \left(  U  \right) = det \left( F \right) Tr \left( F^{-1} \cdot U \right)
\end{eqnarray*}
We are denoting the coordinate $X^{K}\left( X \right)$ by $X^{K}$.\\
Let $\left(T, S, X^{I} , Y^{J} , Y^{J}_{I} \right)$ be the induced local coordinates from the canonical coordinates $\left( T, X^{I} \right)$ of $\mathbb{R}\times \mathbb{R}^{3}$ in $\Phi \left( \mathcal{V}_{0} \right)$ (see example \ref{8} in the appendix I). Then, the evolution equation is given by
\begin{eqnarray*}
 0 &=& \lambda \left(t \right) \  \dot{\mu}\left( t \right) \   \widehat{W}   \left( r \left( j^{1}_{X,X} \phi \right) , J\left( j^{1}_{X,X} \phi \right)\right) \\
&&+\  2 \  \dfrac{\partial \widehat{W}}{\partial r_{\vert j_{X,X}^{1}\phi}}   g \left( X\right)\left( F^{j}_{L}   e^{L} \left( X  \right) ,F^{j}_{L} \Theta^{L}_{M}\left(X \right) e^{M} \left( X  \right) \right)\\
&& +\   det \left(F \right)  \dfrac{\partial \widehat{W}}{\partial J_{\vert j_{X,X}^{1}\phi}}  Tr \left( \Theta^{j}_{I}\left(X\right)\right),
\end{eqnarray*}

\noindent{for all $\left( t,s ,j_{X,X}^{1} \phi\right) \in \Phi \left( \mathcal{V}_{0}  \right)$. Then, the evolution equation is satisfied if, and only if,

\begin{eqnarray*}
 \lambda \left(t \right) &\dot{\mu}&\left( t \right) \   \widehat{W}   \left( r \left( j^{1}_{X,X} \phi \right) , J\left( j^{1}_{X,X} \phi \right)\right)  =\\
 &=&-2 \dfrac{\partial \widehat{W}}{\partial r_{\vert j_{X,X}^{1}\phi}}   g \left( X\right)\left( F^{j}_{L}   e^{L} \left( X  \right) ,F^{j}_{L} \Theta^{L}_{M}\left(X \right) e^{M} \left( X  \right) \right)\\
&=& - det \left(F \right)  \dfrac{\partial \widehat{W}}{\partial J_{\vert j_{X,X}^{1}\phi}}  Tr \left( \Theta^{j}_{I}\left(X\right)\right)
\end{eqnarray*}

Let us now consider two different cases:}
\begin{itemize}
    \item[\textbf{(1)}] \textit{There exists an instant $t$, such that $ \dot{\mu}\left( t \right) \neq 0$.}\\\\
    \noindent{Then, by differentiability, $\dot{\mu} \neq 0$ in an open interval containing $t$. So, the above equation may be considered as an equality in such a way that, on the left side we have a function of $t$, but the other side is distinct of zero and depend on $F$. Notice that the side depending on $F$ cannot be constant. So, the equation cannot fulfilled and, therefore, the dimension of the space of solution around $t$ is $zero$. Then, in this case, the $X-$body-material leaf at $t$ is an interval.}

    \item[\textbf{(2)}] \textit{$\mu$ is constant.}\\\\
    \noindent{Hence, $ \dot{\mu} = 0$, and the equation is satisfied if, and only if,
    }
    \begin{footnotesize}
    \begin{equation}
 - 2  \dfrac{\partial \widehat{W}}{\partial r_{\vert j_{X,X}^{1}\phi}}   g \left( X\right)\left( F^{j}_{L}   e^{L} \left( X  \right) ,F^{j}_{L} \Theta^{L}_{M}\left(X \right) e^{M} \left( X  \right) \right) =     det \left(F \right)  \dfrac{\partial \widehat{W}}{\partial J_{\vert j_{X,X}^{1}\phi}}  Tr \left( \Theta^{j}_{I}\left(X\right)\right)
    \end{equation}
        \end{footnotesize}
However, it is obvious that the zero matrix satisfies this equation. Then, the dimension of the space of solutions $\lambda$, satisfying the evolution equation is $1$. 
\end{itemize}
Therefore, we may extract the following conclusions:
\begin{itemize}
    \item The evolution is a smooth process of remodeling if, and only if, the material stiffness coefficient remains constant in time, 
    \item The evolution is a smooth process of aging if, and only if, the material stiffness coefficient strictly monotonic.
    
\end{itemize}

\section*{Appendix I: Groupoids}\label{appendixsdas}
\addcontentsline{toc}{section}{Appendix I: Groupoids}

Let us now introduce the notion of \textit{groupoid}. A groupoid may be thought as a natural generalization of the concept of group, and it was introduced by Brandt in 1926 \cite{HBU}. Roughly speaking, it is determined by two sets; a set of ``\textit{points}'' $M$ and a set of ``\textit{arrows}'' $\Gamma$, in such a way that each arrow joins two points of $M$. There is a \textit{composition} partially defined in $\Gamma$, such that two arrows may be composed if the starting point of one is equal to the final point of the other, i.e.,

\begin{center}
\includegraphics[width=6cm]{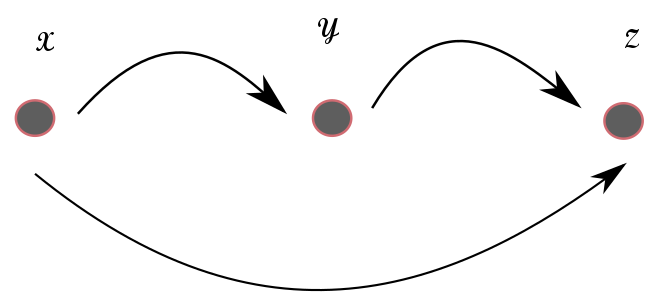}
\end{center}

\begin{definition}
\rm
Let $ M$ be a set. A \textit{groupoid} over $M$ is given by a set $\Gamma$ provided with the maps $\alpha,\beta : \Gamma \rightarrow M$ (\textit{source map} and \textit{target map} respectively), $\epsilon: M \rightarrow \Gamma$ (\textit{section of identities}), $i: \Gamma \rightarrow \Gamma$ (\textit{inversion map}) and $\cdot : \Gamma_{\left(2\right)} \rightarrow \Gamma$ (\textit{composition law}) where for each $k \in \mathbb{N}$, $\Gamma_{\left(k\right)}$ is given by $k$ points $ \left(g_{1}, \hdots , g_{k}\right) \in \Gamma \times \stackrel{k)}{\ldots} \times \Gamma $ such that $\alpha\left(g_{i}\right)=\beta\left(g_{i+1}\right)$ for $i=1, \hdots , k -1$. It satisfy the following properties:\\
\begin{itemize}
\item[(1)] $\alpha$ and $\beta$ are surjective and for each $\left(g,h\right) \in \Gamma_{\left(2\right)}$,
$$ \alpha\left(g \cdot h \right)= \alpha\left(h\right), \ \ \ \beta\left(g \cdot h \right) = \beta\left(g\right).$$
\item[(2)] Associative law, i.e.,
$$ g \cdot \left(h \cdot k\right) = \left(g \cdot h \right) \cdot k, \ \forall \left(g,h,k\right) \in \Gamma_{\left(3\right)}.$$
\item[(3)] For all $ g \in \Gamma$,
$$ g \cdot \epsilon \left( \alpha\left(g\right)\right) = g = \epsilon \left(\beta \left(g\right)\right)\cdot g .$$

\item[(4)] For each $g \in \Gamma$,
$$i\left(g\right) \cdot g = \epsilon \left(\alpha\left(g\right)\right) , \ \ \ g \cdot i\left(g\right) = \epsilon \left(\beta\left(g\right)\right).$$

\end{itemize}
These maps are called \textit{structure maps}. The usual notation for a groupoid is $ \Gamma \rightrightarrows M$.
\end{definition}

Topological or differential structures may be defined on a groupoid.  In particular, we are interested differentiable structure called \textit{Lie groupoids} which was firtly presented by Ehresmann in a series of articles \cite{CELC,CELP,CES,CEC} and redefined in \cite{JPRA} by Pradines.
\begin{definition}
\rm
A \textit{Lie groupoid} is a groupoid $\Gamma \rightrightarrows M$ such that $\Gamma$ is a smooth manifold, $M$ is a smooth manifold and the structure maps are smooth. Furthermore, the source and the target map are submersions.
\end{definition}
A \textit{subgroupoid} of a groupoid $\Gamma \rightrightarrows M$ is a groupoid $\Gamma' \rightrightarrows M'$ such that $M' \subseteq M$, $\Gamma' \subseteq \Gamma$ and the structure maps of $\Gamma'$ are the restriction to $\Gamma'$ of the structure maps of $\Gamma$. A \textit{Lie subgroupoid} of a Lie groupoid $\Gamma \rightrightarrows M$ is a Lie groupoid $\Gamma' \rightrightarrows M'$ which is a subgroupoid of $\Gamma \rightrightarrows M$ such that $\Gamma' $ and $M'$ are submanifolds of $\Gamma$ and $M$ respectively. $\Gamma' \rightrightarrows M'$ is said to be a \textit{reduced Lie subgroupoid} if it is transitive and $M'=M$.
\begin{definition}\label{58}
\rm
Let $\Gamma \rightrightarrows M$ be a Lie groupoid with $\alpha$ and $\beta$ the source map and target map, respectively. For each $x \in M$, the set $\beta^{-1}\left(x\right)$ is called the $\beta-$\textit{fibre at $x$} and denoted by $\Gamma^{x}$. Analogously, the set $\alpha^{-1}\left(x\right)$ is called the $\alpha-$\textit{fibre at $x$} and denoted by $\Gamma_{x}$. Furthermore, the group,
$$\Gamma^{x}_{x}= \beta^{-1}\left(x\right) \cap \alpha^{-1}\left(x\right),$$
is called the \textit{isotropy group of} $\Gamma$ at $x$. The set
$$\mathcal{O}\left(x\right) = \beta\left(\alpha^{-1}\left(x\right)\right) = \alpha\left(\beta^{-1}\left( x\right)\right),$$
is called the \textit{orbit} of $x$, or \textit{the orbit} of $\Gamma$ through $x$.
\end{definition}

\noindent{Notice that, it satisfies that the $\beta-$fibres and the $\alpha-$fibres are closed submanifolds of $\Gamma$ \cite{KMG}}.
\begin{definition}\label{9}
\rm
Let $\Gamma \rightrightarrows M$ be a groupoid. We may define the \textit{left translation on $g \in \Gamma$} as the map $L_{g} : \Gamma^{\alpha\left(g\right)} \rightarrow \Gamma^{\beta\left(g\right)}$, given by
$$ h \mapsto  g \cdot h .$$
We may define the right translation on $g$, $R_{g} : \Gamma_{\beta\left(g\right)} \rightarrow \Gamma_{ \alpha \left(g\right)}$ analogously. 
\end{definition}

For any $ g \in \Gamma $, the left (resp. right) translation on $g$, $L_{g}$ (resp. $R_{g}$), is a diffeomorphism with inverse $L_{g^{-1}}$ (resp. $R_{g^{-1}}$).\\

Of course, there is much more to learn about theory of (Lie) groupoids. However, we would like to focused on the strictly necessary to understand the paper. For a detailed introduction to this topic, we recommend the most relevant reference on groupoids \cite{KMG}. In \cite{EPSBOOK} and \cite{WEINSGROUP} we can also find more intuitive views. The book \cite{JNM} (in Spanish) is also recommendable as a rigurous introduction to groupoids.\\

\begin{example}\label{8}
\rm
Let be a body-time manifold $\mathcal{C}$. Consider $\Phi\left( \mathcal{V}\right)$ the set of all linear isomorphisms between fibres of  vertical subbundle $\mathcal{V}$ of $\mathcal{C}$.\\
For each three material points $X,Y,Z \in \mathcal{B}$, any three instants $t,s,r \in \mathbb{R}$ and each two $1-$jets, $j_{X,Y}^{1} \psi$ and $j_{Y,Z}^{1} \phi$, of a (local) diffeomorphism, the structure maps of the groupoid $\Phi\left( \mathcal{V}\right)$ are given by,
\begin{itemize}
\item[(i)] $\alpha\left(t,s, j_{X,Y}^{1} \psi\right) = \left( t,X \right)$
\item[(ii)] $\beta\left(t,sj_{X,Y}^{1} \psi\right) = \left( s,Y \right)$
\item[(iii)] $ \left(s,r, j_{Y,Z}^{1} \phi\right)  \cdot \left(t,s, j_{X,Y}^{1} \psi\right) = \left( t,r, j_{X,Z}^{1} \left( \phi \circ \psi \right)\right)$
\end{itemize}
This groupoid $\Phi\left( \mathcal{V}\right) \rightrightarrows \mathcal{C}$ is called the \textit{vertical frame groupoid of $\mathcal{C}$}.\\
Let $\left( x^{i}\right)$ and $\left( y^{j}\right)$ be two local coordinates defined on the open subsets of $\mathcal{B}$, $\mathcal{U}$ and $\mathcal{W}$ respectively. Then, we may define a system of local coordinates of $\Phi \left( \mathcal{V}\right)$ as follows,
\begin{equation}\label{17.second2313}
\Phi \left(  \mathcal{V}_{\mathcal{U}, \mathcal{W}}\right) : \left(t,s,x^{i} , y^{j}, y^{j}_{i}\right),
\end{equation}
where, for each $ \left( t , s , j_{X,Y}^{1}\phi \right) \in \Phi \left(  \mathcal{V}_{\mathcal{U}, \mathcal{W}}\right) $
\begin{itemize}
\item $t \left( t , s , j_{X,Y}^{1}\phi \right) = t$.
\item $s\left( t , s , j_{X,Y}^{1}\phi \right) = s.$
\item $x^{i} \left( t , s , j_{X,Y}^{1}\phi \right) = x^{i} \left(X\right)$.
\item $y^{j} \left( t , s , j_{X,Y}^{1}\phi \right) = y^{j} \left( Y\right)$.
\item $y^{j}_{i}\left( t , s , j_{X,Y}^{1}\phi \right)  = \dfrac{\partial \left(y^{j}\circ \phi\right)}{\partial x^{i}_{| X} }$.
\end{itemize}
where $\Phi \left(  \mathcal{V}_{\mathcal{U}, \mathcal{W}} \right)$ is given by the any triple $\left( t , s , j_{X,Y}^{1}\phi \right)$ such that $X \in \mathcal{U}$ and $Y \in \mathcal{W}$. These local coordinates turns $\Phi\left( \mathcal{V}\right) \rightrightarrows \mathcal{C}$ into a Lie groupoid.

\end{example}

\section*{Acknowledgments}
M. de Leon and V. M. Jiménez acknowledge the partial finantial support from MICINN Grant PID2019-106715GB-C21 and the ICMAT Severo Ochoa project CEX2019-000904-S. 

\bibliographystyle{plain}

\bibliography{Library}

\end{document}